\documentclass[reprint,amsmath,amssymb,aps]{revtex4-1}
\usepackage{graphicx}
\usepackage{dcolumn}
\usepackage{bm}
\usepackage{color}
\usepackage[margin=2cm]{geometry}
\usepackage{epstopdf}

\begin{document}

\title{Equations of state and stability of $\text{MgSiO}_3$ perovskite and post-perovskite phases from quantum Monte Carlo simulations}
\author{Yangzheng Lin$^1$}
\author{R. E. Cohen$^{1,2}$} \email[E-mail: ]{rcohen@carnegiescience.edu}
\author{Stephen Stackhouse$^{3,5}$}
\author{Kevin P. Driver$^{3}$}
\author{Burkhard Militzer$^{3,4}$}
\author{Luke Shulenburger$^{6}$}
\author{Jeongnim Kim$^{7}$}
\affiliation{$^1$Geophysical Laboratory, Carnegie Institution of Washington, 5251 Broad Branch Road, NW, Washington DC 20015, USA}
\affiliation{$^2$Department of Earth Science, University College London, UK}
\affiliation{$^3$Departments of Earth and Planetary Science and $^4$of Astronomy, University of California, Berkeley, CA 94720, USA}
\affiliation{$^5$School of Earth and Environment, University of Leeds, Leeds, UK}
\affiliation{$^6$Sandia National Laboratories, Albuquerque, NM, USA}
\affiliation{$^7$Oak Ridge National Laboratory, Oak Ridge, TN, USA}

\begin{abstract}
We have performed quantum  Monte Carlo (QMC) simulations and density functional theory (DFT) calculations to study the equations of state of $\text{MgSiO}_3$ perovskite (Pv) and post-perovskite (PPv), up to the pressure and temperature conditions of the base of Earth's lower mantle. The ground state energies were derived using QMC and the temperature dependent Helmholtz free energies were calculated within the quasi-harmonic approximation and density functional perturbation theory. The equations of state for both phases of $\text{MgSiO}_3$ agree well with experiments, and better than those from generalized gradient approximation (GGA) calculations. The Pv-PPv phase boundary calculated from our QMC equations of states is also consistent with experiments, and better than previous LDA calculations. We discuss the implications for double crossing of the Pv-PPv boundary in the Earth.
\end{abstract}


\maketitle


\section{Introduction}

The accurate description of electronic correlation effects is one of the main challenges in theoretical condensed matter physics. Quantum Monte Carlo (QMC) \cite{Ceperley80,Perdew81,Ceperley86,Foulkes01,Needs10} simulations can describe these correlation effects while maintaining a high computational efficiency \cite{Ceperley06}. A number of recent studies demonstrate the growing ability of the QMC method to accurately describe ground state properties of complex solids \cite{Alfe04,Alfe05,Kolorenc08,Driver10,Esler10,Abbasnejad12,Shulenburger13,Kim14}. This large basis of previous work provides the motivation to apply QMC calculations to solid silicate perovskite (Pv) and post-perovskite (PPv) (MgSiO$_3$) in order to derive equations of state that are more accurate than those that have been previously obtained with density function theory (DFT).

The Pv-PPv phase transition is particularly important because Pv is the dominant phase in Earth's lower mantle \cite{Knittle87}. Pv was the only known phase under lower mantle conditions until a phase transition to PPv at a pressure of 125 GPa and temperature of 2500 K was discovered in 2004 \cite{Murakami04,Oganov04}. The post-perovskite phase is believed to exist in Earth's thin, core-mantle boundary layer, known as $\text{D}^{\prime\prime}$. The discovery of MgSiO$_3$ PPv has attracted considerable attention as it offers a possible explanation for many of the unusual properties of the $\text{D}^{\prime\prime}$ layer, such as the inhomogeneous seismic discontinuity observed a few hundred kilometers above the core-mantle boundary, anomalous seismic anisotropy, and ultra-low velocity zones \cite{Sidorin99,Garnero00,Murakami04,Oganov04,Wookey05,Mao06}. Some quantitative estimates of these anomalies were made by Oganov et al. \cite{Oganov04}

Many computations \cite{Cohen93,Karki00,Karki01,Oganov01,Oganov04,Tsuchiya04,Iitaka04,Tsuchiya05,Cohen05,Liu12} based on DFT \cite{Hohenberg64,Kohn65} have reported the equations of state of Pv and PPv. However, DFT results are dependent on the choice of exchange-correlation functional \cite{Hamann96,Cohen06,Driver10} since the exact exchange-correlation functional is unknown. Generally, DFT with the local density approximation (LDA) provides a good P-V relationship for MgSiO$_3$ perovskite \cite{Cohen93,Karki00}, but underestimates the Pv-PPv transition pressures \cite{Oganov04,Tsuchiya04}. In contrast, whereas DFT with the generalized gradient approximation (GGA) provides a better prediction of the Pv-PPv transition pressure, it overestimates the zero pressure lattice volume in the equation of state \cite{Oganov04}. The $\sim$10 GPa difference between the LDA and GGA predictions of the phase transition pressure \cite{Tsuchiya04} makes a difference in depth of about 150 km, according to the preliminary reference Earth model (PREM) \cite{Dziewonski81}. The discrepancy among DFT calculations, although relatively small for many applications, are significant with regard to geophysical modelling. The position of the Pv to PPv boundary is crucial for interpreting seismic data from the base of the mantle, to understand if this transition is sufficient to explain most lower mantle heterogeneity, or if there must also be partial melt, compositional heterogeneity, etc. In particular, double crossing of the Pv-PPv boundary, \cite{hernlund2005,hernlund2007} or more generally, through the two-phase region, \cite{hernlund2010} may give indication of temperature and compositional variations, which are crucial for interpreting the seismological data, \cite{lay2006} and as input into geodynamic modelling \cite{tackley2007}. 

\section{Computational Methods}

\subsection{Quantum Monte Carlo}

A rigorous discussion of QMC methods has been reported in previous publications \cite{Foulkes01,Needs10}. Here, we briefly outline the main choices we make within the methodology. We employ two types of QMC sequentially in order to extract the ground state properties of a system. The first is known as variational Monte Carlo (VMC), in which a fixed-form trial many-body is constructed by multiplying a single-particle Slater determinant by a Jastrow correlation factor:
\begin{equation}
\Psi_{\text{T}}\left(R\right)=D^{\uparrow}D^{\downarrow}e^J,
\label{eq:TWFunc}
\end{equation}
The up- and down-spin Slater determinants, $D^{\uparrow}$ and $D^{\downarrow}$, are obtained from DFT calculations. The Slater determinant fixes the nodal surface of the calculation, which is used in the so-called fixed-node approximation. The Jastrow factor, $J$, is the exponential of a sum of parameterized one-body and two-body terms that are a function of particle separation, and satisfies the cusp condition. The Jastrow parameters are optimized by minimizing a combination of variance of the VMC energy and the energy itself \cite{Umrigar05}.

VMC by itself is generally not accurate enough due to the fixed form of the trial wavefunction. In a second method, diffusion Monte Carlo (DMC), a statistical representation of the wavefunction is evolved according to a version of the Schr\"odinger equation which has been transformed to an imaginary time diffusion equation. The statistical wavefunction, constructed from the optimized trial VMC wavefunction, is evolved in imaginary time until it exponentially decays to the ground state. The DMC method is very efficient at projecting out the ground state as all higher energy states are exponentially damped:
\begin{equation}
\Psi_{\textbf{DMC}}=\lim_{\Delta t\rightarrow 0}\prod_{j=1}^{N}e^{-H\Delta t}\Psi_{\text{T}},
\label{eq:DMCWFunc}
\end{equation}
where $H$ is the Hamilitonian, $\Delta t$ is step size used for imaginary time propagation, and $N$ corresponds to the number of projections. In both VMC and DMC, the space of electron configurations is simultaneously explored with an ensemble of independent configurations, which follow a random walk. In DMC, the walk is guided by an importance-sampled wave-function for efficiency. Once configurations have equilibrated, averages of their energies can be accumulated and analyzed statistically. This allows QMC methods to be massively parallelized in computations.

Diffusion Monte Carlo would be an exact, stochastic, solution to the Schr\"odinger equation except for the sign problem, which is controlled by using fixed many-body wave-function nodes. The nodal surface comes from the trial wave function, which in our case is from a single Slater determinant of the Kohn-Sham orbitals from a converged DFT computation. At least there is a variational principle, so we can say that our result is an upper bound on the total energy. In MgSiO$_3$, an insulating, closed shell system, we expect this approximation to be very good, and in general to be independent of structure or compression, so even if there is a small shift in the total energy from inaxct nodes, it should be very close in each structure studied. The agreement we discuss below with experiment is post-hoc evidence of this. The second approximation is related to the use of pseudopotentials, discussed more below. This would be no worse than the use of pseudopotentials in DFT, except that there is an additional ``locality approximation" which can make results sensitive to the pseudopotentials used. This approximation can be controlled and tested, and again should give similar errors to each structure for the MgSiO$_3$ system. Again, the evidence is that this is usually a very small error.

\subsection{Pseudopotentials}
While great care is taken to generate pseudopotentials, they are constructed within the mean field treatment rather than a many-body framework. A recent paper which systematically applied diffusion Monte Carlo to calculate properties of solids concluded that the determining factor on the accuracy of the method was the fidelity of the pseudopotentials used \cite{Shulenburger13}. This conclusion echoed results of earlier studies on geophysics with QMC which found the effects of the pseudopotential approximation to be large \cite{Alfe05}. One possible strategy to mitigate this error is to perform all electron calculations as was done in recent work on boron nitride \cite{Esler10}. This approach is however impractical for this study due to the extreme computational demands posed by all electron calculations within QMC for heavier elements.  As an alternative, we have endeavored to test the pseudopotentials used in this paper as rigorously as possible in an attempt to validate their use in DMC.

Our testing methodology has three parts. First we have tested the Mg and O pseudopotentials by comparing LAPW calculations performed with the ELK code \cite{Dewhurstxx} to pseudopotential calculations performed with quantum espresso \cite{Giannozzi09}. In calculations of the equilibrium lattice constant and bulk modulus we find excellent agreement with the all electron results giving 4.219 {\AA } and 156.6 GPa and the pseudopotential results giving 4.206 {\AA}  and 159.9 GPa.  While this test is not conclusive in terms of stating that the pseudopotential will be accurate for QMC calculations, it does preclude corrections of the type applied in earlier work \cite{Alfe05}.

The second test we applied was to calculate the electron affinity and ionization potential for each of the pseudopotentials used and to compare the results to experiment as was shown to be useful in a recent paper on Ca$_2$CuO$_3$ \cite{Foyevtsova14}. Here we use the Slater-Jastrow form for the trial wavefunction with a single Slater determinant. The spin state for the neutral atom, anion and cation are determined from spin polarized DFT calculations and are kept fixed in the QMC. While this is not likely to produce highly accurate electron affinities or ionization potentials, large errors greater than approximately 0.1 eV would be a cause of concern. In this case however, the pseudopotentials prove to be rather accurate as shown in Table \ref{tab:ionization}.

\begin{figure}[h]
\includegraphics[width=2.4in,angle=-90]{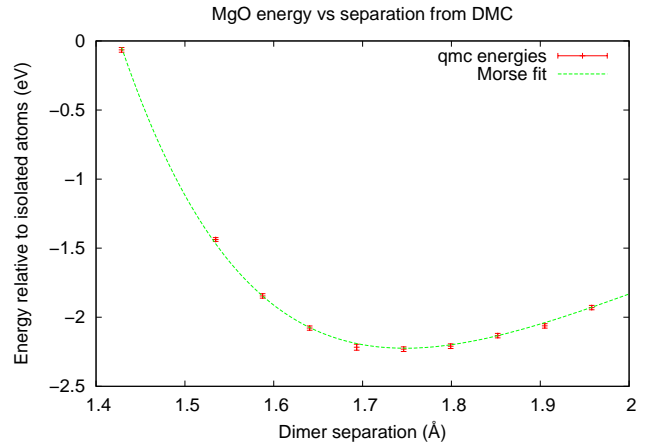}
\caption{Energy vs separation curve for MgO calculated using DMC.}
\label{fig:MgO-binding}
\end{figure}

The third test we applied was to calculate binding curves of the molecules MgO, O$_2$ and SiO. These energy vs. separation curves were then fitted to a Morse potential and the resulting atomization energies, bond lengths and vibrational frequencies are compared to experiment where possible. From the example of MgO case in Fig. \ref{fig:MgO-binding}, we could see the fitting of Morse potential to the QMC energy is pretty good. It should again be noted that an attempt to make these calculations as accurate as possible would use a more sophisticated wavefunction containing for instance a multideterminant expansion \cite{Morales12}. However, the performance of the single Slater-Jastrow trial wavefunction is highly relevant as this is the form used in calculation of the properties of the solid phases.  In this case we find excellent agreement with experiments (Table \ref{tab:molecule}), leading us to conclude that these pseudopotentials are accurate for use in calculating the perovskite to post-perovskite phase transition pressure.

\begin{widetext}
\begin{table}
\caption{Electron affinity (EA) and ionization potential (IP) from QMC and experiment given in eV.  The electron affinity of Si and O is taken from Ref. \onlinecite{Chaibi10}, while the negative electron affinity of Mg is from Ref. \onlinecite{Andersen04}. The ionization potentials are taken from Ref. \onlinecite{Lide03}.}
\label{tab:ionization}
\begin{tabular}{@{}lcccc@{}}
\hline
 & QMC EA & Expt EA & QMC IP & Expt IP \\ \hline
Mg   & unbound & unbound & 7.591$\pm$0.013  & 7.64624 \\
O  & 1.372$\pm$0.013 & 1.4611134 & 13.681$\pm$0.026 & 13.61806 \\
Si  & 1.430$\pm$0.013 & 1.3896210 & 8.228$\pm$0.012 & 8.15169 \\ 
\hline
\end{tabular}
\end{table}

\begin{table}[h]
\caption{Equilibrium bond length ($r_e$) in angstroms, vibrational frequency ($\omega_0''$) in cm$^{-1}$ and atomization energy ($D_0$) in eV calculated with QMC and fit to a Morse potential compared to experiments.  Note that the vibrational frequency of SiO is difficult to obtain experimentally due to issues in isolating SiO at low temperatures.
\label{tab:molecule}}
\begin{tabular}{@{}lcccccc@{}}
\hline
 & QMC $r_e$ & Expt $r_e$ & QMC $\omega_0''$ & Expt $\omega_0''$ & QMC $D_0$ & Expt $D_0$ \\
\hline
MgO   & 1.7519$\pm$0.0018 & 1.749 \cite{Huber79} & 782.3$\pm$32.5 & 785.2183$\pm$0.0006\cite{Irikura07} & 2.14$\pm$0.02 & 2.56$\pm$0.22\cite{Operti89}\\
O$_2$  & 1.1978$\pm$0.0007 & 1.208 \cite{Huber79} & 1480.5$\pm$11.4 & 1580.161$\pm$0.009 \cite{Irikura07} & 4.89$\pm$0.02 & 5.117\cite{Chase85} \\
SiO  & 1.5097$\pm$0.0007 & 1.51 \cite{Huber79} & 788.9$\pm$22.6 & 1241.54388$\pm$0.00007\cite{Irikura07} & 7.98$\pm$0.42 & 8.24\cite{Chase85}\\ 
\hline
\end{tabular}
\end{table}
\end{widetext}
\newpage

It is possible to make high quality pseudo potentials for correlated systems \cite{trail2013}. However, such pseudopotentials would not be applicable to the DFT computations we use to generate our trial functions, so one would have to use different pseudopotentials to generate the trial functions. Put in this way, one could say that the problem is not with pseudo potentials per se, but in ones that are useable for generation of trial functions, or that the ultimate problem is with the trial functions.  In principle the trial functions could be parametetrized and optimised variationally, \cite{neuscamman2012}, or backflow \cite{rios2006}, or other nodal variations could be used, but such computations have not yet been possible for complex solids such as we study here.

\subsection{Quasi Harmonic Phonon free energies}

Whereas the static crystal energy can be obtained by DFT \cite{Hohenberg64,Kohn65} or QMC \cite{Ceperley80,Perdew81,Ceperley86} calculations, it is currently intractable to calculate the phonon frequencies from quantum Monte Carlo simulations. Therefore, in our results we combine static QMC energies with vibrational energies from density functional perturbation theory (DFPT) calculations. The accuracy of QMC static energy plus DFPT vibrational energies has been shown to be an improvement over using DFT plus DFPT for the silica phases \cite{Driver10}. Once the Helmholtz free energies are obtained for several lattice volumes at various temperatures, the temperature dependent equation of state and other thermodynamic properties of interest are determined.

The Helmholtz free energy is a function of lattice volume and temperature $F\left(V,T\right)$. Using the Vinet equation of states \cite{Vinet87,Cohen00}, the Helmholtz free energy is
\begin{equation}
\begin{split}
F=&F_0+\frac{4K_0V_0}{\left(K'_0-1\right)^2}-\frac{2K_0V_0}{\left(K'_0-1\right)^2}\\
&\left\{5+3K'_0\left[\left(\frac{V}{V_0}\right)^{1/3}-1\right]-3\left(\frac{V}{V_0}\right)^{1/3}\right\}\\
&\exp{\left\{-\frac{3}{2}\left(K'_0-1\right)\left[\left(\frac{V}{V_0}\right)^{1/3}-1\right]\right\}}
\end{split}
\label{eq:FofV}
\end{equation}
where $F_0$, $V_0$, $K_0$ and $K'_0$ are the Helmholtz free energy, lattice volume, bulk modulus and its pressure derivative respectively, under zero pressure. Within the quasi-harmonic approximation (QHA), the Helmholtz free energy is given by \cite{Born54,Wallance72},
\begin{equation}
\begin{split}
F=&E+TS=E_{\text{Static}}+\frac{1}{2}\sum_{\textbf{k},i}\hbar\omega_{\textbf{k},i}\\
&+k_{\text{B}}T\sum_{\textbf{k},i}\ln{\left[1-\exp{\left(\frac{-\hbar\omega_{\textbf{k},i}}{k_{\text{B}}T}\right)}\right]}
\end{split}
\label{eq:FofT}
\end{equation}
where $E$ is the internal energy, $S$ is the entropy, $E_{\text{Static}}$ is the static energy, $\hbar$ is the Planck constant/$2\pi$, $\omega_{\textbf{k},i}$ is the angular frequency of a phonon with wave vector $\textbf{k}$ in the $i$-th band, $k_{\text{B}}$ is the Boltzmann constant and $T$ is the absolute temperature. In the quasi-harmonic approximation, $E_{\text{Static}}$ and $\omega_{\textbf{k},i}$ are independent of $T$ and are determined only by the atomic positions and lattice parameters at zero temperature.

\subsection{Computational details}

The pseudopotentials used for all DFT, DFPT and QMC calculations in this work were generated with the OPIUM code \cite{Opiumxx} using the WC exchange correlation functional \cite{Cohen06}. The core radii of the pseudopotentials are as follows: 1.2(1s), 1.2(2s) for magnesium; 1.3(1s), 1.3(2s) for oxygen; and 1.7(1s), 1.7(2s), 1.7(2p) for silicon.

Our DFT equations of state are constructed from the energies of seven different volumes in both the Pv and PPv phases. We used the plane-wave pseudopotential DFT code, PWSCF \cite{Giannozzi09}, to relax the atomic position, obtain the static DFT energy, and extract the single-particle orbitals for the QMC wavefunction at each volume. The seven volumes correspond to constant pressure simulations at  -20, -10, 0, 50, 100, 150, and 200 GPa. These calculations used the Wu-Cohen (WC) \cite{Cohen06} exchange correlation approximation, a plane-wave energy cutoff of 300 Ry and Monkhorst-Pack k-point meshes of $6 \times 6 \times 6$ and $12 \times 6 \times 6$ for the 20 atoms unit cell of perovskite and post-pervoskite respectively, which converged the total energy to tenths of milli-Ry/$\text{MgSiO}_3$ accuracy. At each volume above, the phonon frequencies and temperature dependent vibrational energies were calculated with ABINIT \cite{Gonze09} using density functional perturbation theory within the quasi-harmonic approximation. These calculations used q-point meshes of $4 \times 4 \times 4$ for the 20 atom unit cell of perovskite and $4 \times 4 \times 2$ for the 10 atom unit cell of post-perovskite, which ensured calculated phonon free energies were converged to tenths of milli-Ry/$\text{MgSiO}_3$.

The accuracy of our QMC calculations is determined by three classes of approximations that are necessary for computational efficiency of fermionic calculations: finite simulation cell size effects, pseudopotentials, and the fixed node approximation \cite{Drummond08}. For accurate QMC results, one must reduce the error introduced by these approximations such that the end result is converged. The effectivity of pseudopotentials in QMC calculations was checked in the previous section. Here, we discuss how the other approximations are mitigated.

In any simulation we are forced to simulate a true solid with a simulation cell that is subject to periodic boundary conditions. Finite-size errors arise from both one-body effects due to discrete k-point sampling of the Brillouin zone and two-body effects from spurious electron correlation in the periodic cells. We minimize the one-body errors by using twist averaged boundary conditions. We average over eight twists, allowing us to improve our sampling of the Brillouin zone. The two-body errors are minimized by using the Model Periodic Coulomb (MPC) interaction \cite{Drummond08,Fraser96,Williamson97}, which corrects the potential energy for the spurious correlation effects. We then use the scheme of Chiesa et al. \cite{Chiesa06} to correct the kinetic energy two-body effects. While applying these techniques, we then perform our calculations in three different supercell sizes of 40, 80, and 120 atoms, and we fit an extrapolation to infinite cell size.

The final approximation we will discuss is that of the nodal surface to handle the fermion sign problem. QMC samples a positive definite probability function constructed from a antisymmetric wavefunction which has positive and negative regions. Unchecked, sampling the probability in this way will lead to a bosonic ground sate as positive and negative contributions cancel out and the odd-parity solution becomes swamped in statistical noise. In order to circumvent this problem, absorbing barriers are placed between all nodal pockets in configuration space. This can only be done if the nodes are fixed to a known location at the start of DMC (we use the nodes from DFT), which is called the fixed-node approximation. The size of the fixed-node error is generally assumed to be small, and, for small systems, can be checked with backflow optimization of the single-particle orbital coordinates, but this is too computationally expensive for the systems studied here.

In order to ensure electron correlation was treated uniformly across all of our calculations, we fixed the Jastrow parameters in DMC simulations for all volumes and supercell sizes to values obtained from optimizing the Jastrow for the smallest volume and supercell size. In addition, for computational efficiency, a b-spline basis set is used to represent the single particle orbitals centered on a grid of points. The b-spline basis set provides an order-$N$ speed up in the calculation, where $N$ is the number of atoms, but doubles the memory requirement relative to an analytic, plane-waves basis. The mesh size of this grid is decreased until the total energy is converged to tenths of milli-Ry/$\text{MgSiO}_3$. We use a b-spline mesh factor of 0.8. 

For the wavefunction optimization part of our calculations, a combination of energy and variance minimization was used in a series of twenty optimizations in which the VMC total energy was determined to a one-sigma statistical accuracy of 0.05 eV/$\text{MgSiO}_3$ and the fluctuation among the VMC energies after each optimization became less than 0.1 eV/$\text{MgSiO}_3$. The Jastrow factor which gave both lowest total energy and smallest variance was chosen for use in the subsequent DMC simulations.

A typical DMC simulation used 300-400 electron configurations and collected statistics over 25,000 Monte Carlo steps. The first 5000 steps were used to equilibrate the simulation. The total energy of each supercell was obtained by averaging the energies of the remaining 20,000 steps over 8 twists. The standard error $\delta$ of the total energy was obtained by $\delta = \sqrt{\sigma^2/M}$, where $\sigma$ is the energy variance of block samples and $M$ is the uncorrelated samples \cite{Kim12}. The DMC time-step was determined by converging the total energy with respect to changes in the time-step. Our convergence tests found that a time-step of 0.001 $\text{Ha}^{-1}$ is sufficient for 0.05 eV/$\text{MgSiO}_3$ accuracy. The VMC and DMC simulations were preformed with the GPU version of QMCPACK \cite{Kim12,Esler12,QMCPackxx}.

\section{Results and Discussion}

\subsection{Enthalpy and Volume}
\begin{figure}[h]
\includegraphics[width=3.2in]{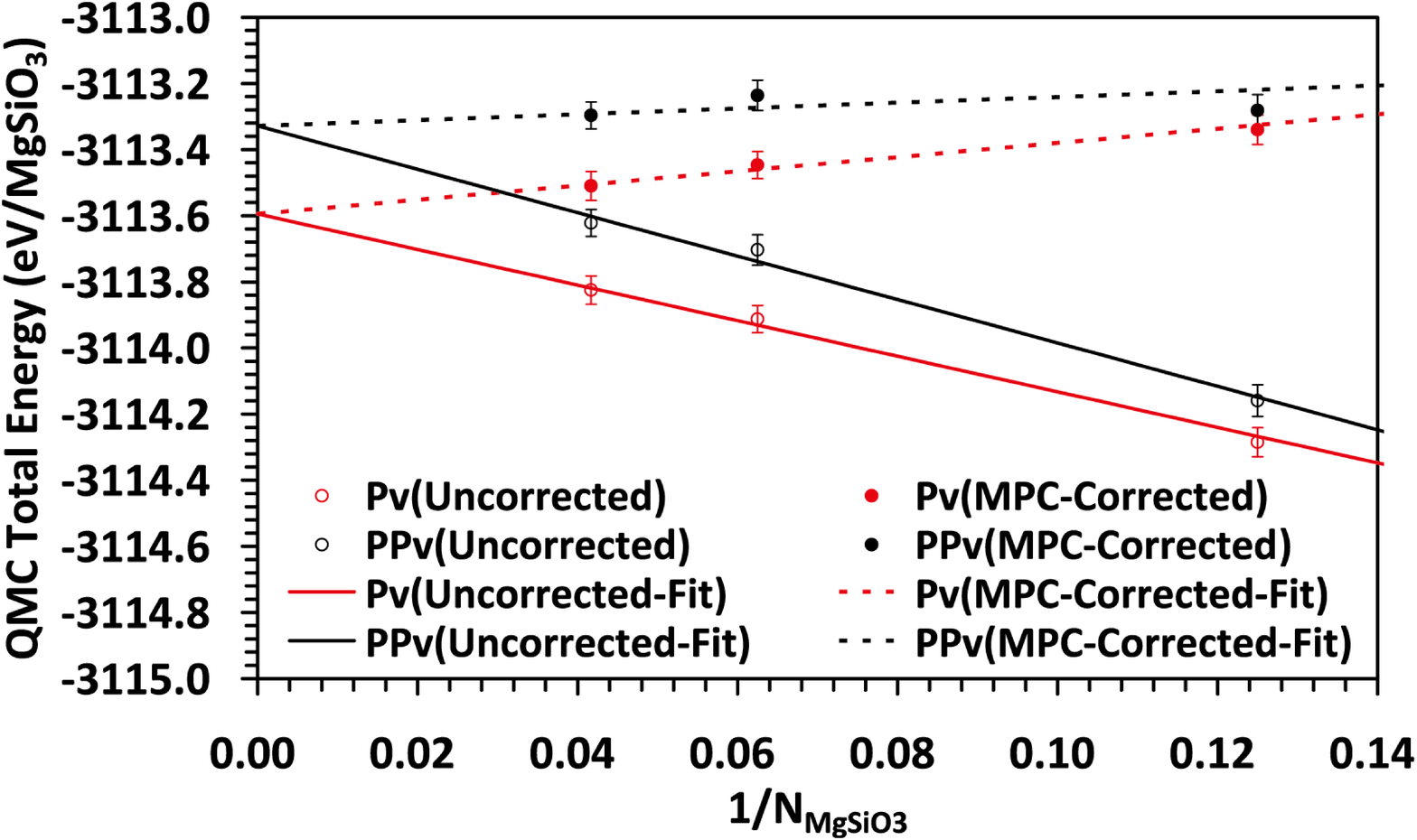}
\caption{The extrapolation to infinite size QMC total energy using the MPC-corrected and uncorrected finite size total energies. All the total energies are the 8 twists averaged results. $N_{\text{MgSiO}_3}$ is the number of formula units in the supercell used in the QMC simulation. Here the volume is 41.19 \AA$^3$/MgSiO$_3$ for Pv and 41.15 \AA$^3$/MgSiO$_3$ for PPv.}
\label{fig:Figures02}
\end{figure}

In our simulations, both the MPC corrected and uncorrected QMC total energies as a function of simulation cell size are quite linear (Fig. \ref{fig:Figures02}). All QMC results hereafter are from DMC simulations unless stated otherwise. Two linear equations were used to fit the MPC-corrected and uncorrected total energies synchronously. The equations are $E_{N}^{\text{cor}}=E_{\inf}^{\text{cor}}+k^{\text{cor}}/N$ for MPC corrected energies and $E_{N}^{\text{un-cor}}=E_{\inf}^{\text{un-cor}}+k^{\text{un-cor}}/N$ for uncorrected energies, where $k$ is the slope and $N$ is the number of MgSiO$_3$ formula. $E_{\inf}^{\text{cor}}$ and $E_{\inf}^{\text{un-cor}}$ were kept equal to each other during the fitting process using the least squares method and they are our final infinite size QMC total energy, which is the static energy in Eq. \ref{eq:FofT}. The error of the infinite size energy was taken to be the same as that of the largest supercell size case.

\begin{figure}[h]
\includegraphics[width=3.2in]{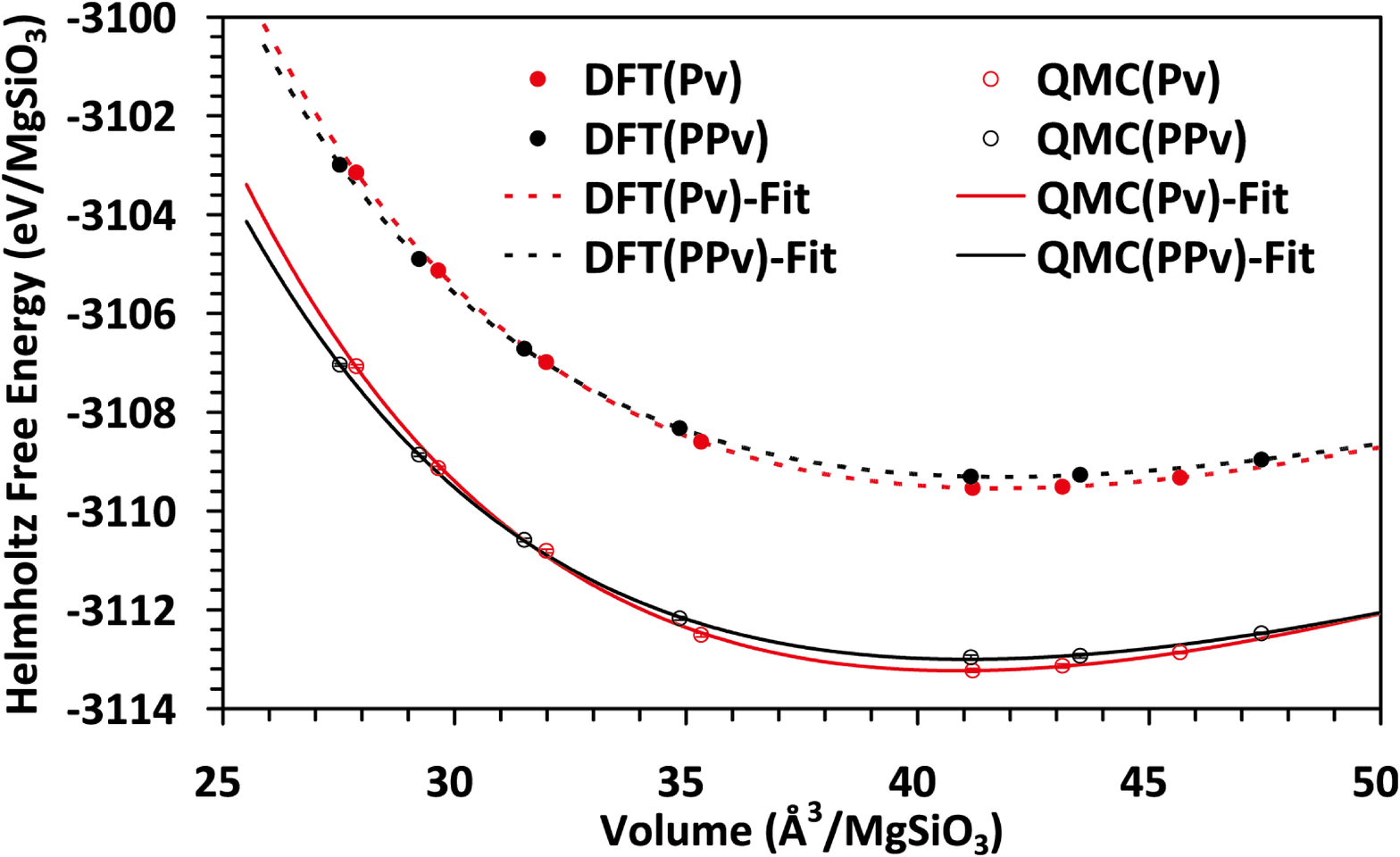}
\caption{QMC and DFT Helmholtz free energies of $\text{MgSiO}_3$ Pv and PPv at 300 K, the error bars of QMC results are covered by the symbols. The dashed and solid lines are the fit of the Vinet equation of state.}
\label{fig:Figures03}
\end{figure}

At a given temperature, the Vinet equation of state (Eq. \ref{eq:FofV}) was used to fit the Helmholtz free energies as a function of volume. The fittings both for DFT and QMC calculations are quite good as shown in Fig. \ref{fig:Figures03}. The predicted equilibrium volume, bulk modulus and its pressure derivative from QMC simulations at 300 K are all in good agreement with experimental results (Table \ref{tab:Table03}) both for Pv and PPv. This indicates QMC is better than GGA because all the equilibrium volumes predicted by this and previous GGA calculations are larger than experimental data, and the bulk modulus predicted by GGA are smaller than experiments. With the increase of temperature, the bulk modulus decreases while it's derivative on pressure increases (Fig. \ref{fig:FiguresM1}).

\begin{widetext}

\begin{table}[t]
\centering
\caption{Equation of state parameters of pervoskite and post-perovskite. The uncertainties of QMC thermodynamic quantities were propagated from the errors of QMC static energies via a linearized Taylor expansion.}
\label{tab:Table03}
\begin{tabular}{@{}llllll@{}}
\hline
& $F_0$(eV/MgSiO$_3$)& $V_0$(\AA$^3$/MgSiO$_3$) & $K_0$(GPa) & $K'_0$ & \\
\hline
Pv & & &  & &  \\
 &-3113.61(3) & 40.36(8) & 270(14) & 3.9(4) & QMC, Static, this work  \\
 &-3113.23(3) & 40.88(10) & 258(15) & 4.0(4) & QMC, 300 K, this work  \\
 &-3109.90 & 41.20 & 239.4 & 4.1 & GGA (WC), Static, this work \\
 &-3109.54 & 41.79 & 226.6 & 4.2 & GGA (WC), 300 K, this work \\
 & --      & 40.5  & 259   & 4.01 & LDA, Static \cite{Karki01} \\
 & --      & 41.03 & 247   & 3.97 & LDA, 300K \cite{Karki00,Karki01} \\
 &-- & 40.2 & 266 & 4.2 & LDA, Static \cite{Cohen93} \\
 &-- & 40.85 & 259.8 & 4.14 & LDA, 300 K \cite{Oganov04} \\
 &-- & 41.85 & 230.1 & 4.06 & GGA, 300 K \cite{Oganov04} \\
 &-- & 41.02 & 248 & 3.9 & LDA, 300 K \cite{Tsuchiya04} \\
 &-- & 41.03 & 246 & 4.0 & LDA, 300 K \cite{Tsuchiya05} \\
 &-- & 38.53 & 271 & 3.74 & LDA, Static \cite{Cohen05} \\
 &-- & 40.78 & 232 & 3.86 & GGA, Static \cite{Cohen05} \\
 &-- & 40.58-40.83 & 246-272 & 3.65-4.00 & Exp. \cite{Yagi78,Knittle87,Kudoh87,Mao89,Ross90,Mao91,Wang94,Fiquet00,Shieh06} \\
PPv & & &  & &  \\
 &-3113.38(3) & 40.51(8) & 232(9) & 4.1(3) & QMC, Static, this work  \\
 &-3113.00(3) & 41.08(9) & 221(10) & 4.2(3) & QMC, 300 K, this work  \\
 &-3109.67 & 41.19 & 205.0 & 4.6 & GGA (WC), Static, this work \\
 &-3109.31 & 41.85 & 192.3 & 4.7 & GGA (WC), 300 K, this work \\
 &-- & 40.73 & 231.9 & 4.43 & LDA, 300 K \cite{Oganov04} \\
 &-- & 41.9 & 200.0 & 4.54 & GGA, 300 K \cite{Oganov04} \\
 &-- & 40.95 & 222 & 4.2 & LDA, 300 K \cite{Tsuchiya04} \\
 &-- & 40.95 & 215.9 & 4.41 & LDA, 300 K \cite{Tsuchiya05} \\
 &-- & 38.4 & 243 & 4.05 & LDA, Static \cite{Cohen05} \\
 &-- & 40.8 & 203 & 4.19 & GGA, Static \cite{Cohen05} \\
 &-- & 40.85 & 209 & 4.4 & GGA (PW91), Static \cite{Liu12} \\
 &-- & 40.55-41.23 & 219-248 & 4.0(fixed) & Exp. \cite{Shieh06,Ono06,Guignot07} \\
\hline
\end{tabular}
\end{table}

\begin{figure}[h]
\includegraphics[width=6.4in]{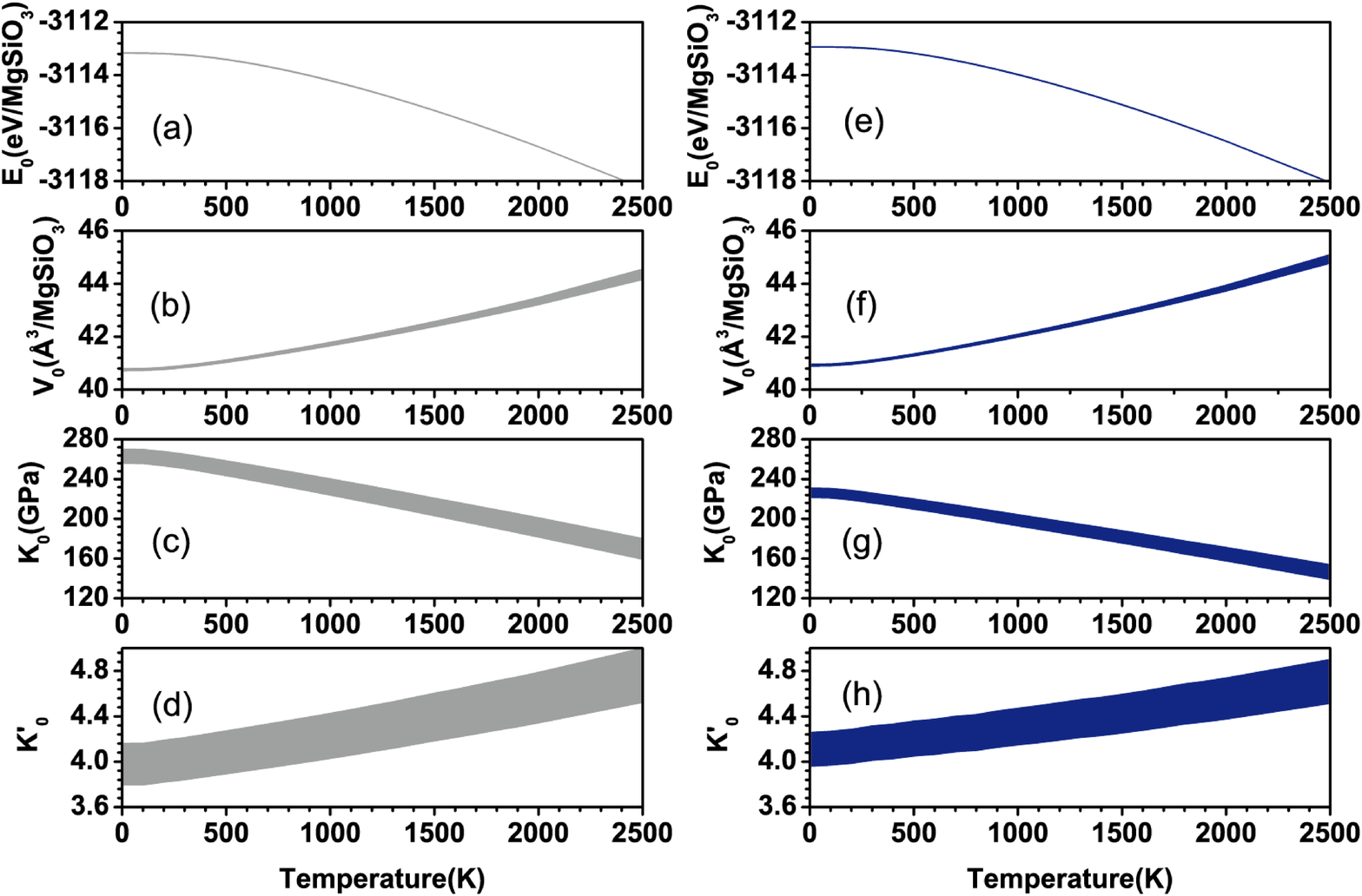}
\caption{Vinet equation of state parameters based on QMC static energies and DFPT vibrational energies as a function of temperature. (a), (b), (c) and (d) are for perovskite. (e), (f), (g) and (h) are for post-perovskite. The shading of the curves indicate one-sigma width of the uncertainties.}
\label{fig:FiguresM1}
\end{figure}

\end{widetext}

\subsection{P-V-T equation of state}
\begin{figure}[htbp]
\includegraphics[width=3.2in]{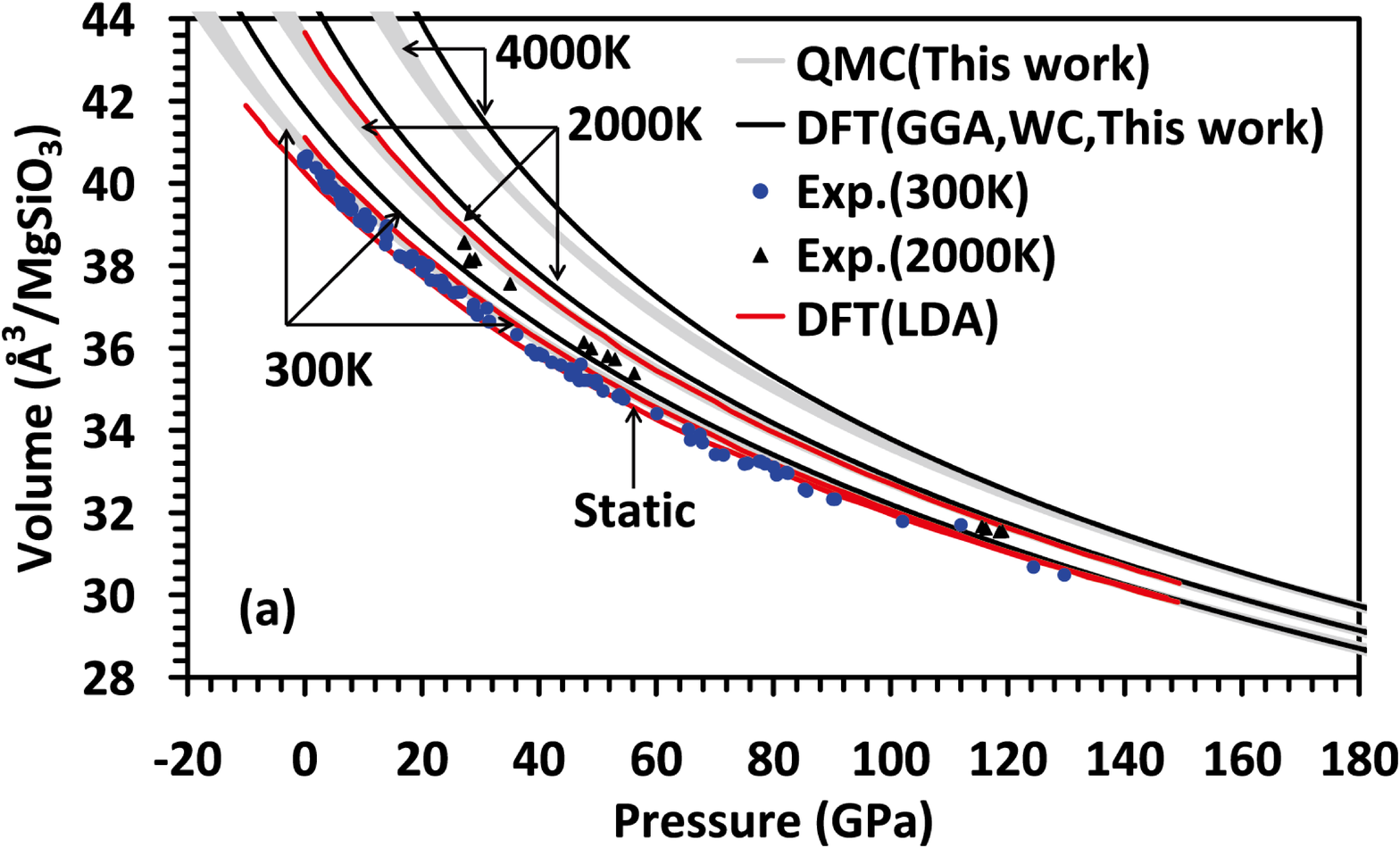}
\includegraphics[width=3.2in]{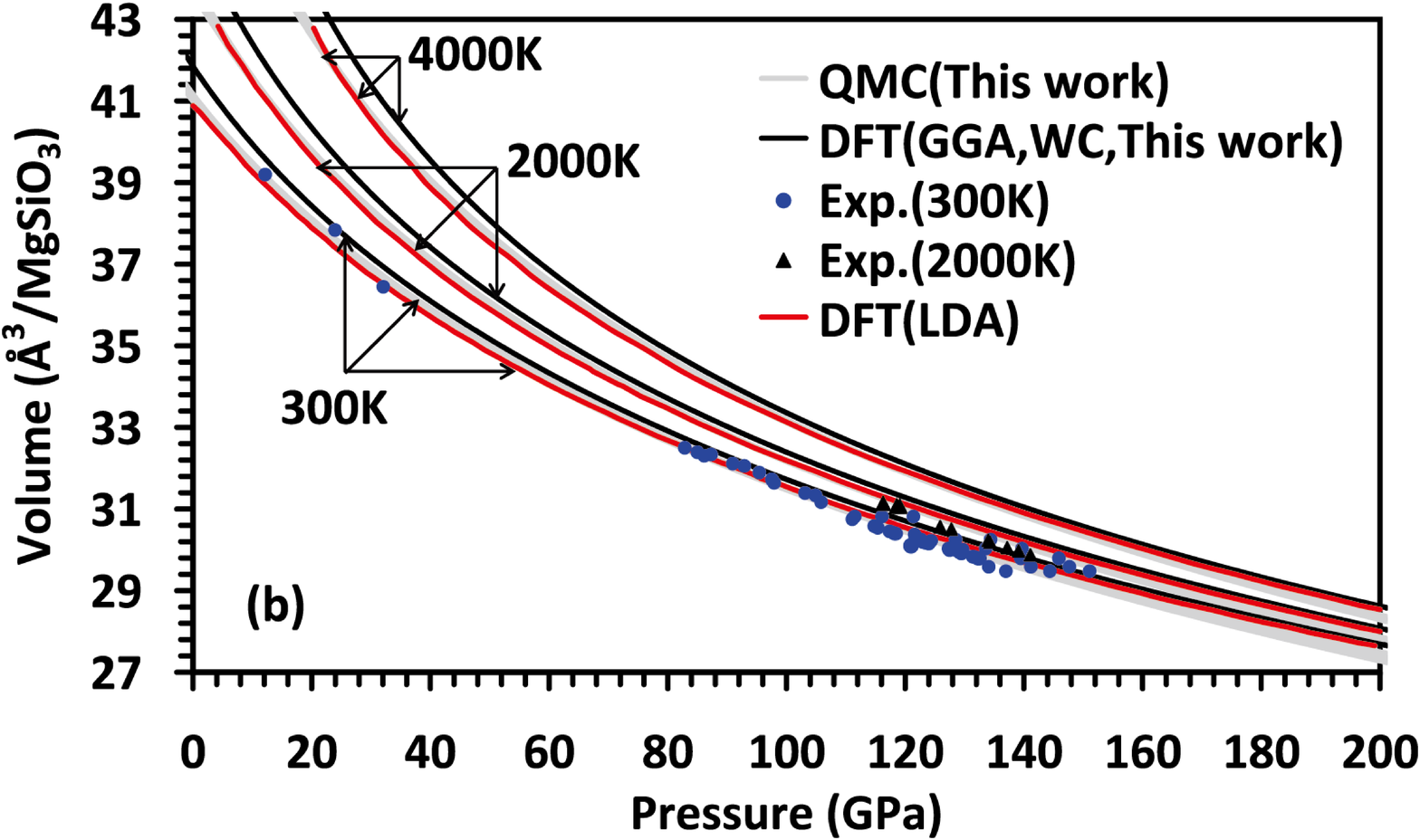}
\caption{Thermal equations of state of (a) perovskite and (b) post-perovskite at three temperatures. The static LDA results of Pv are taken from Ref. \cite{Cohen93}, and the finite temperature LDA results of Pv are taken from Ref. \cite{Karki00}. The LDA results of PPv are taken from Ref. \cite{Tsuchiya05}. The experimental results of Pv are taken from Refs.  \cite{Cohen93,Fiquet00,Shieh06,Kudoh87,Utsumi95,Funamori96,Fiquet98,Saxena99,Komabayashi08,Deng08} at 300 K and from Refs.  \cite{Funamori96,Fiquet98,Komabayashi08} at 2000 K. The experimental results of PPv are taken from Refs.  \cite{Murakami04,Shieh06,Ono06,Guignot07,Komabayashi08} at 300 K and Refs. \cite{Guignot07,Komabayashi08} at 2000 K. The shading width of QMC results indicates two-sigma statistic errors at the given pressure.}
\label{fig:Figures04}
\end{figure}

The thermal equation of state can be calculated by $P=-(\partial F/\partial V)_T$ from the Helmholtz free energy Eq. \ref{eq:FofV}. The comparison between the computed thermal equations of state and previous experimental data for both Pv and PPv phases are figured in Fig. \ref{fig:Figures04}. The shading of the QMC curves in Fig. \ref{fig:Figures04} indicate the width of standard deviation of volume as a function of pressure caused by the statistical errors of QMC energies. These comparisons indicate our QMC simulations and LDA calculations predicted a better P-V-T relationship than GGA calculations for both Pv and PPv. The LDA calculations for PPv were taken from Ref. \cite{Tsuchiya05} where the comparison with experiments was not checked. 

\begin{figure}[h]
\centering{\includegraphics[width=3.2in]{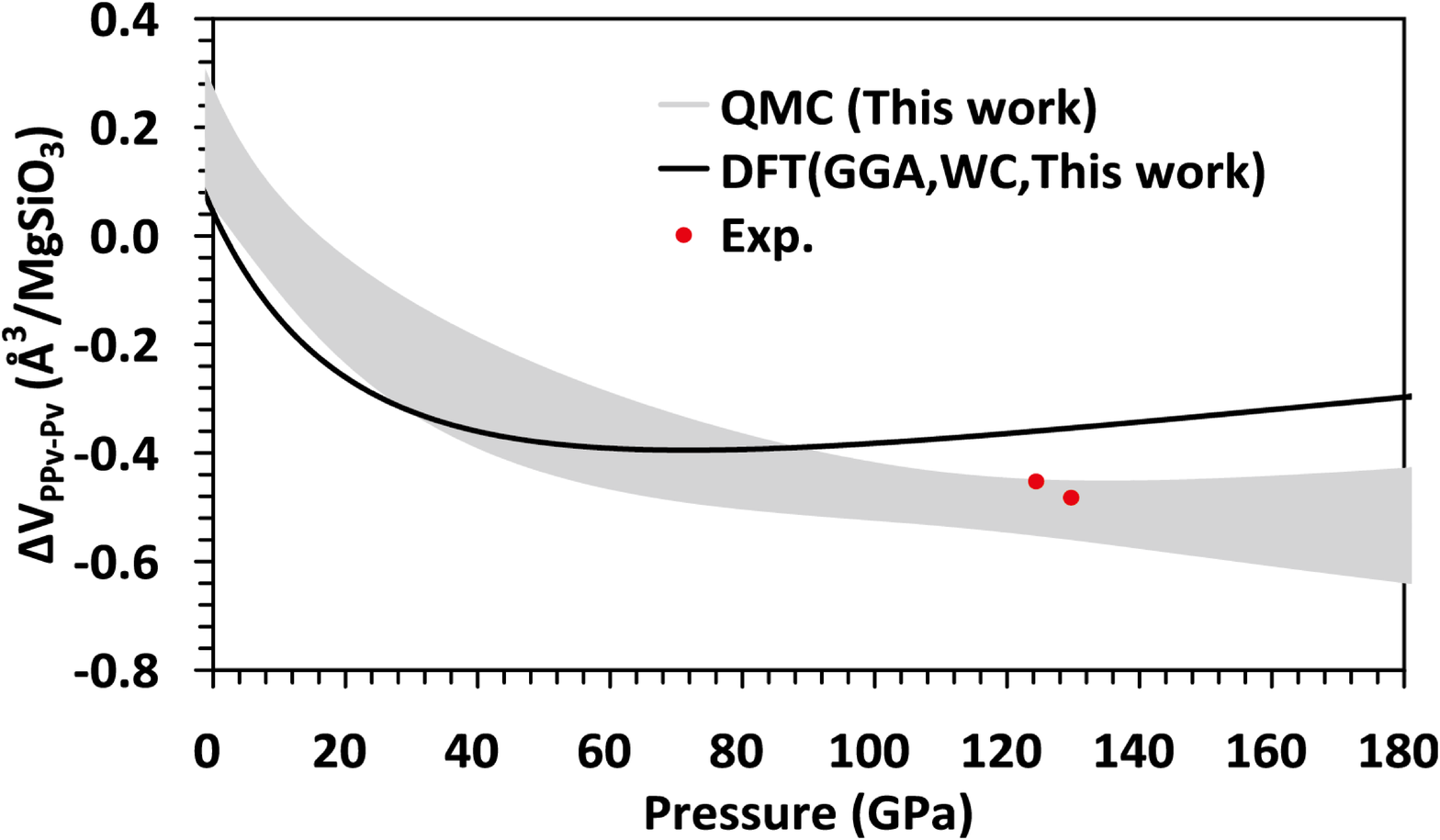}}
\caption{Volume differences between perovskite and post-perovskite phases in our DFT and QMC calculations and their comparison with experiments \cite{Komabayashi08} at 300 K. The shading width indicates two-sigma statistic errors at the given pressure.}
\label{fig:FiguresS1}
\end{figure}

We also calculated the volume differences between MgSiO$_3$ perovskite and post-perovskite phases as a function of pressure and compared them with some available experimental data in Fig. \ref{fig:FiguresS1}. The comparison indicates that our QMC results are in good agreement with experiments \cite{Komabayashi08}. At lower mantle conditions, our Pv-PPv volume difference is much closer to experiment than DFT.

\subsection{Phase boundary}
\begin{figure}[h]
\includegraphics[width=3.2in]{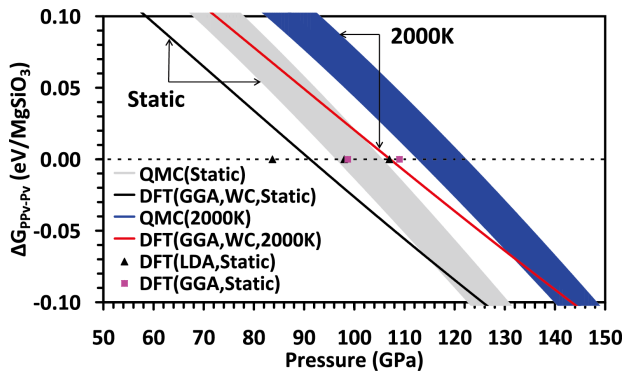}
\caption{Gibbs free energy difference between $\text{MgSiO}_3$ pervoskite and post-perovskite phases in static state and at 2000 K. The shading width of QMC results indicates one-sigma statistic errors at the given pressure. The triangle data of LDA results are from Refs. \cite{Iitaka04,Oganov04,Tsuchiya04,Cohen05} and the square data of GGA results are from Refs. \cite{Oganov04,Tsuchiya04}.}
\label{fig:Figures05}
\end{figure}

In thermodynamics, the Gibbs free energy $G$ is defined as $G=F+PV$. At a fixed temperature, a phase transition occurs when Gibbs free energy of the current phase becomes greater than that of another phase with the change of pressure. Because of the uncertainty of static total energy from QMC simulations, we could only predict a range of transition pressure as shown in Fig. \ref{fig:Figures05}. Due to the fact Gibbs free energy differences between $\text{MgSiO}_3$ Pv and PPv are very small, the range of transition pressure from QMC simulations looks somehow wide. In spite of that, the predicted one-sigma range of transition pressure by QMC simulations still has obvious deviation from that predicted by DFT calculations for $\text{MgSiO}_3$ Pv and PPv phases. In static state, we obtained a Pv-PPv phase transition pressure of 91.2 GPa from GGA results and, at a one sigma intervals, 101.0$\pm$4.6 GPa from QMC results. Again, we see the transition pressure predicted by this DFT calculation is different from previous DFT studies. At 2000 K, we obtained a Pv-PPv phase transition pressure of 107.1 GPa from DFT (GGA) results and, at a one sigma interval, 117.5$\pm$4.8 GPa from QMC results.

\begin{figure}[h]
\includegraphics[width=3.2in]{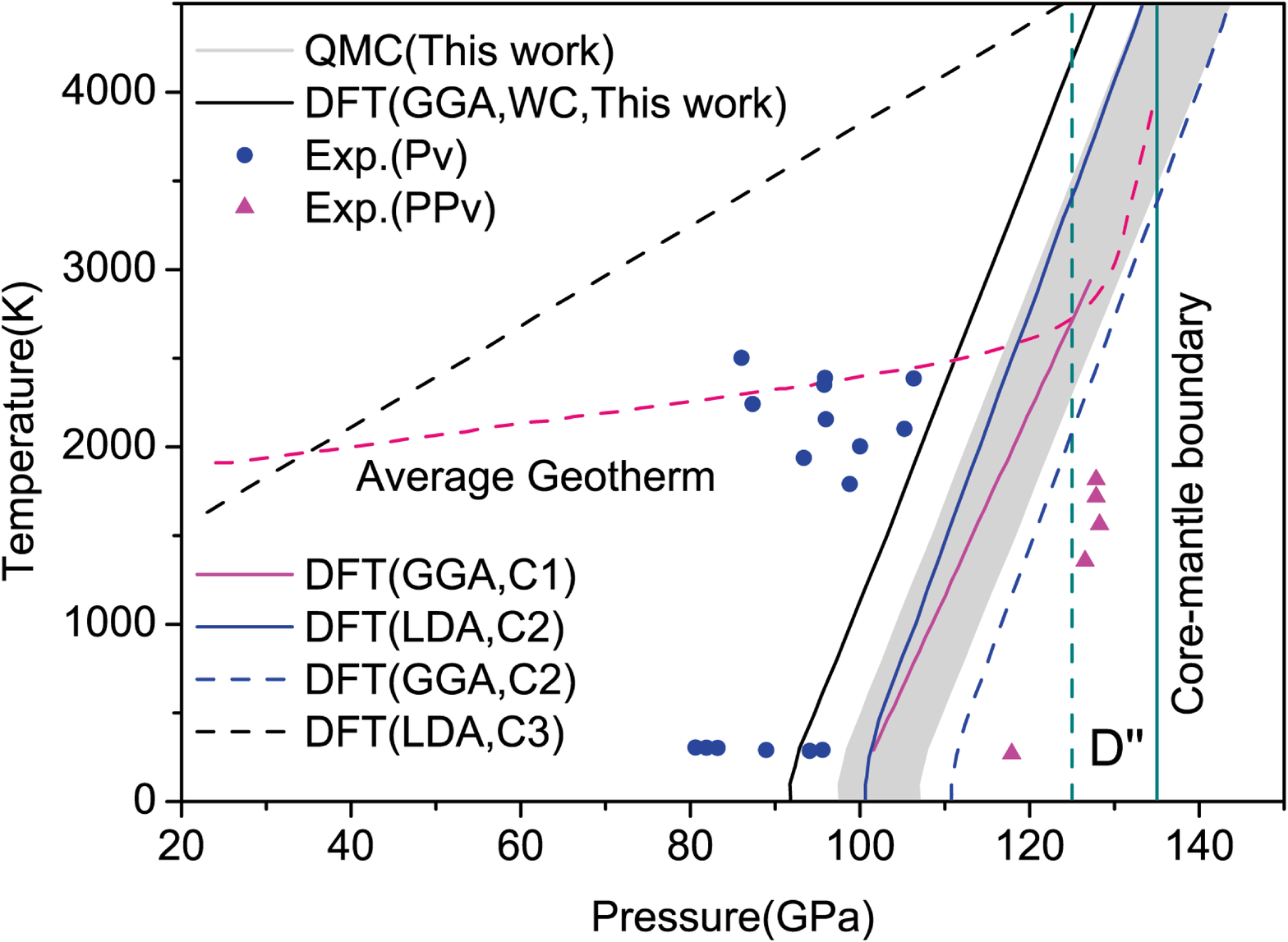}
\caption{Phase diagram of $\text{MgSiO}_3$ under lower mantle pressure conditions. The black solid line and gray shading area showing the one-sigma error band show the Pv-PPv boundaries from our DFT (WC) and QMC computations  respectively. The shading of QMC results indicates one-sigma statistic errors. The DFT C1, C2, and C3 curves are taken from Ref. \cite{Oganov04}, Ref. \cite{Tsuchiya04} and Ref. \cite{Tsuchiya05} respectively. The pressure in the D" layer of the lower mantle falls between the vertical dash line and solid line of the core-mantle boundary. The experimental results are taken from Ref. \cite{Oganov04} The average geotherm is from Ref. \cite{Boehler00}. }
\label{fig:Figures06}
\end{figure}

At any temperature in the range of 0 to 4500 K, the Pv-PPv transition pressure predicted by our DFT computations with the WC exchange-correlation functional \cite{Cohen06} is always smaller than that predicted by our QMC calculations (Fig. \ref{fig:Figures06}), and it falls between the LDA and GGA boundaries predicted by Tsuchiya et al. \cite{Tsuchiya04}. The Pv to PPv transition pressure predicted by QHA within LDA from Ref. \cite{Tsuchiya05} is much lower than that reported in experimental studies and other calculations. The Clapeyron slope is obtained as 8.4$\pm$0.8 MPa K$^{-1}$ based on samples in the QMC phase transition boundary in temperature range of 500$\sim$4500 K. It has been proposed that there is double crossing of the Pv-PPv phase boundary along the geotherm \cite{Hernlund05,Hernlund07}. Our results are consistent with double crossing for \emph{pure} $\text{MgSiO}_3$ (Fig. \ref{fig:Figures06}), but do not require double crossing. However, Fe partitions into PPv, and thus stabilises the PPv phase \cite{Cohen05,Shieh06}. Depending on the exact shape of the two phase region, the double crossing can still give a seismic signature \cite{hernlund2010}. Although some LSDA+U studies for (Mg$_{0.9375}$ Fe$_{0.0625}$)SiO$_3$ suggest that Fe incorporation has only a marginal effect on the high spin Pv to PPv phase transition pressure \cite{Metsue12a,Metsue12b}, they only considered 6.25\% iron substitution. Our GGA+U calculation for pure anti-ferromagnetic FeSiO$_3$ shows that the PPv phase has a static enthalpy 0.14 eV/FeSiO$_3$ lower than the Pv phase at 100 GPa, and at 0 GPa, PPv FeSiO$_3$ still has a static enthalpy 0.10 eV/FeSiO$_3$ lower than Pv FeSiO$_3$. In our MgSiO$_3$ calculations, the vibrational energy of Pv is about 0.09 eV/MgSiO$_3$ lower than that of PPv between 0 and 200 GPa at 4000 K. The vibrational energy difference between the Pv and PPv phases is highly dependent on temperature. Generally, the lower the temperature, the smaller the difference. Iron is thus expected to partition into PPv, and further increase its stability under Earth's lower mantle conditions.

\subsection{Thermodynamic properties}

\begin{figure}[h]
\includegraphics[width=3.2in]{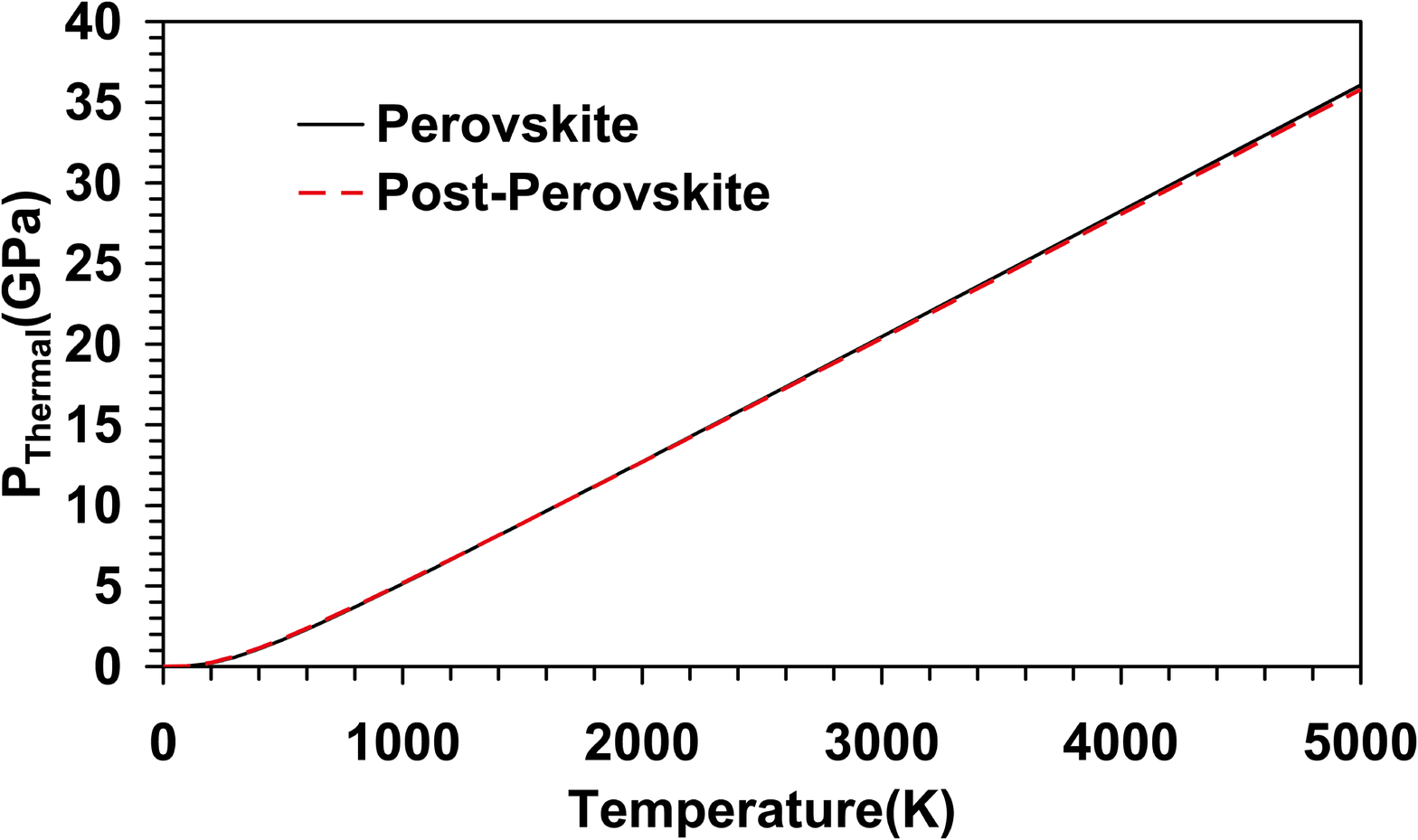}
\caption{The thermal pressures averaged over volume from 27.8 to 44.5 \AA$^3$/MgSiO$_3$ as a function of temperature.}
\label{fig:FiguresM2}
\end{figure}

The thermal pressure is defined as \cite{Jackson96,Cohen01}
\begin{equation}
P_{th}(V,T)=P(V,T)-P(V,0)=-(\partial F_{th}/\partial V)_{T}
\label{eq:Pth}
\end{equation}
where $F_{th}$ is the thermal free energy (the third term in Eq. \ref{eq:FofT}). For either phase of Pv and PPv, the averaged thermal pressure over volume as a function of temperature is quite linear at temperatures larger than 1000 K (Fig. \ref{fig:FiguresM2}). The slopes of the linear parts of thermal pressure curves are 7.74 MPa K$^{-1}$ for Pv and 7.65 MPa K$^{-1}$ for PPv.

\begin{figure}[htbp]
\includegraphics[width=3.2in]{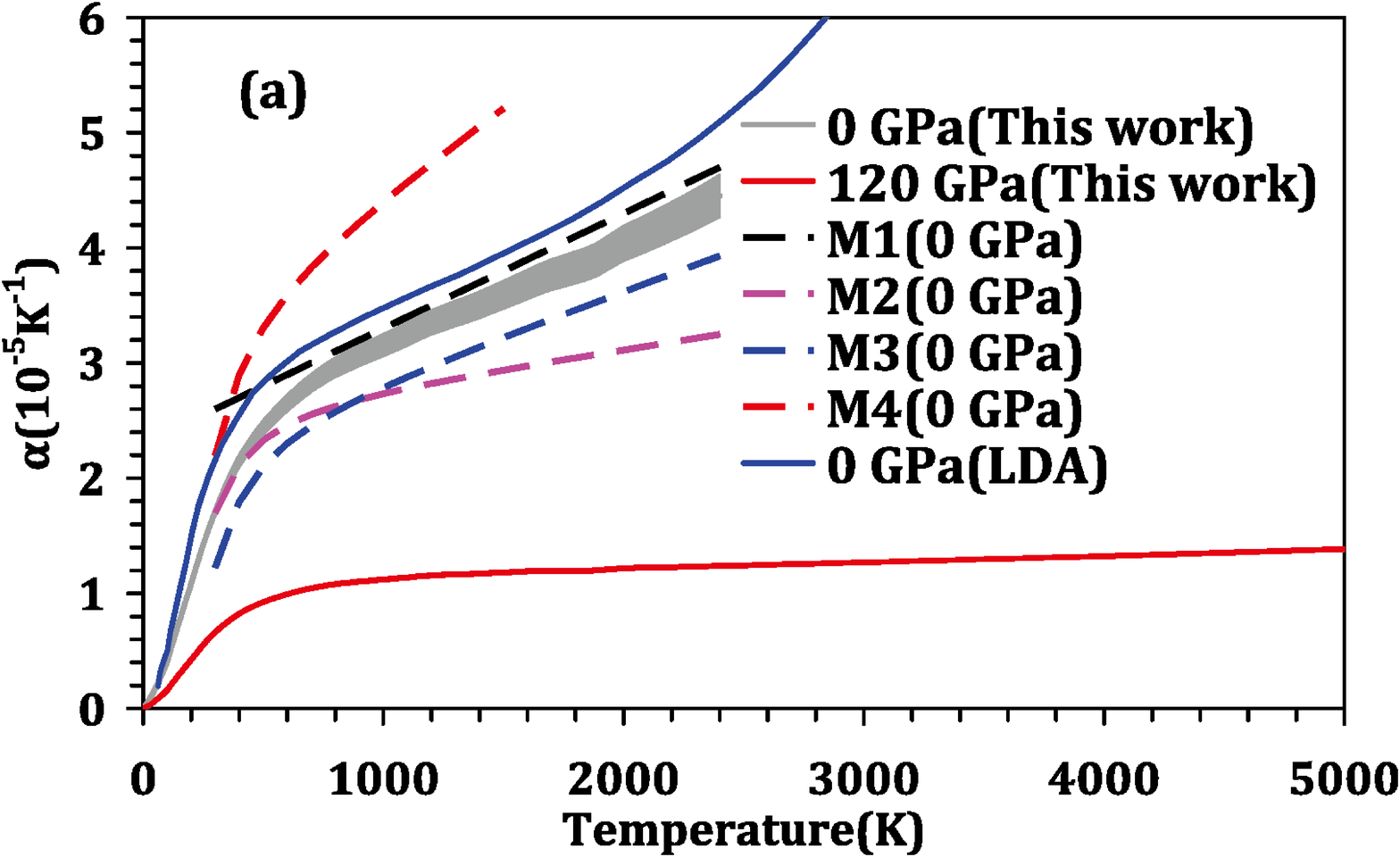}
\includegraphics[width=3.2in]{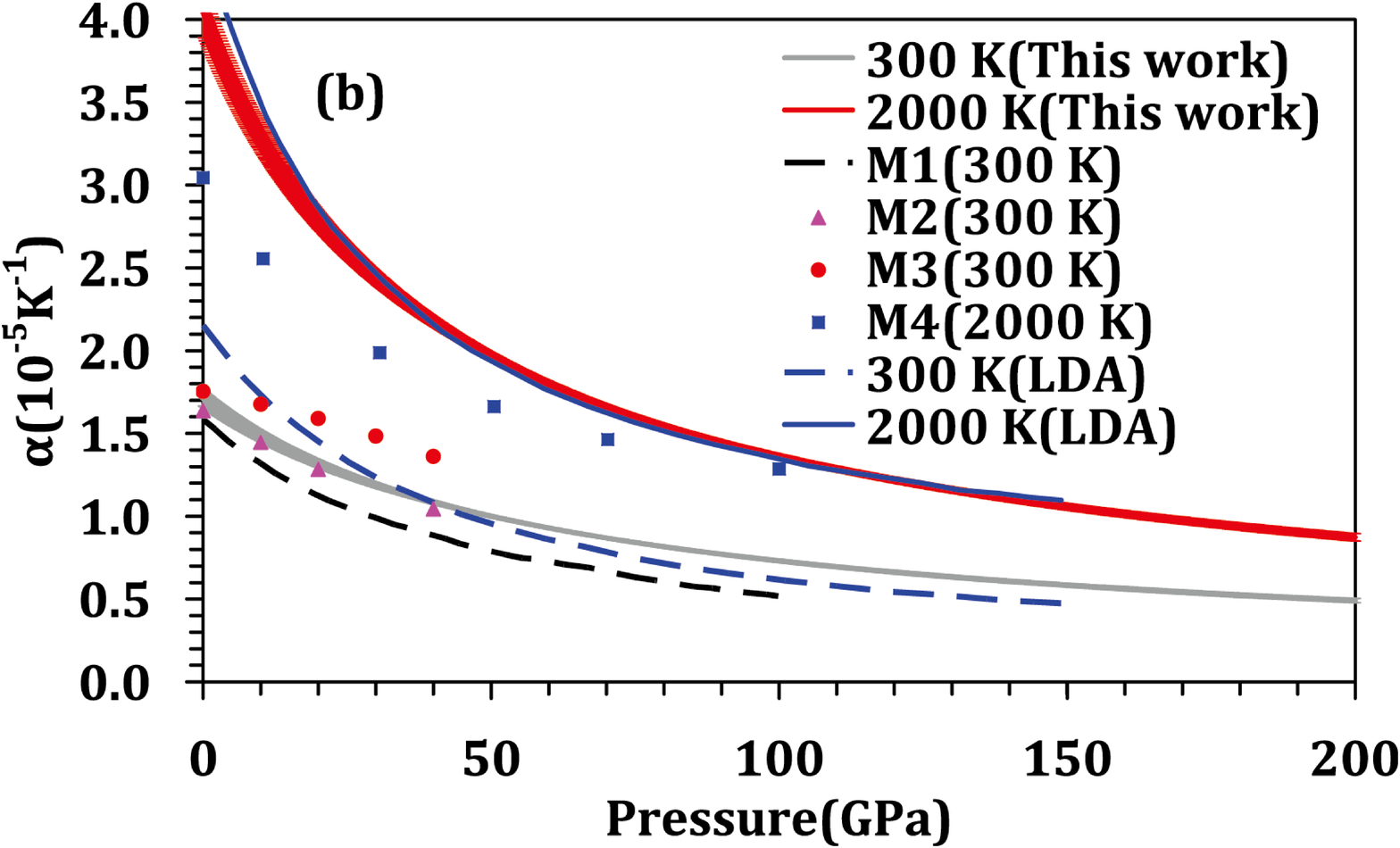}
\includegraphics[width=3.2in]{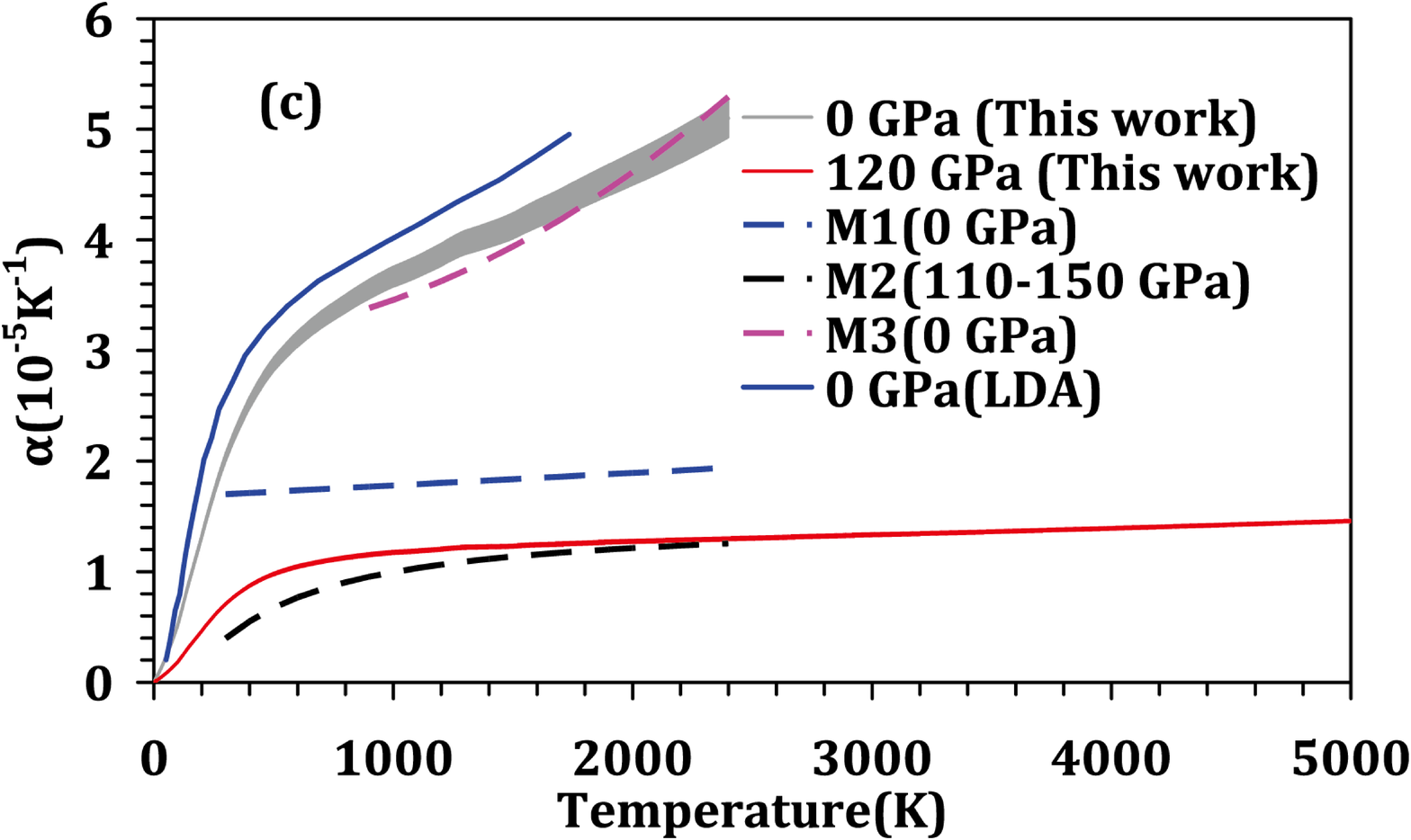}
\includegraphics[width=3.2in]{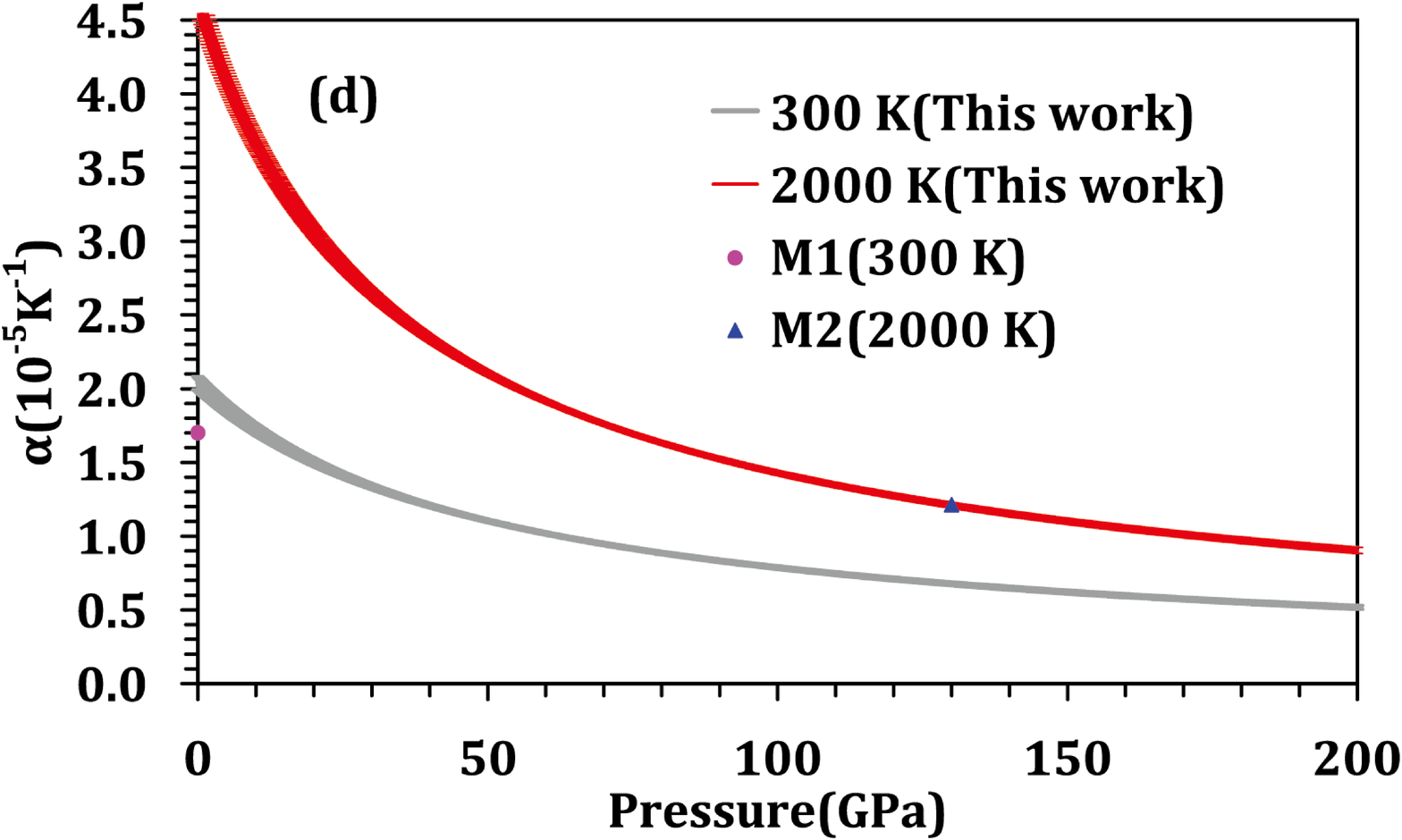}
\caption{Temperature and pressure dependence of the thermal expansion. (a) and (b) are for perovskite. (c) and (d) are for post-perovskite. The shading of the curves indicate one-sigma width of the uncertainties. The references for all the models and DFT calculations are \cite{Katsura09}(a,M1), \cite{Komabayashi08}(a,M2; c,M1; d,M1), \cite{Saxena99}(a,M3; b,M4), \cite{Mao91}(a,M4), \cite{Karki01}{a,LDA}, \cite{Hama98}(b,M1), \cite{Gillet96}(b,M2), \cite{Chopelas96}(b,M3), \cite{Karki00}{b,LDA}, \cite{Guignot07}(c,M2; d,M2), \cite{Ono05}(c,M3), and \cite{Tsuchiya05}{c,LDA}.}
\label{fig:FiguresM3}
\end{figure}

The thermal expansivity $\alpha$ is calculated from thermal pressure as
\begin{equation}
\alpha=(\partial P_{th}/\partial T)_V/K_T
\label{eq:Alpha}
\end{equation}
where $K_T=-V(\partial P/\partial V)_T$ can be obtained from the thermal equation of state. In the next part of this section, the related thermal equation of state is derived based on QMC static energies and DFPT vibrational energies. The obtained thermal expansivities in this work fall in the region of previous models which were derived from experimental data (Fig. \ref{fig:FiguresM3}).  

\begin{figure}[htbp]
\includegraphics[width=3.2in]{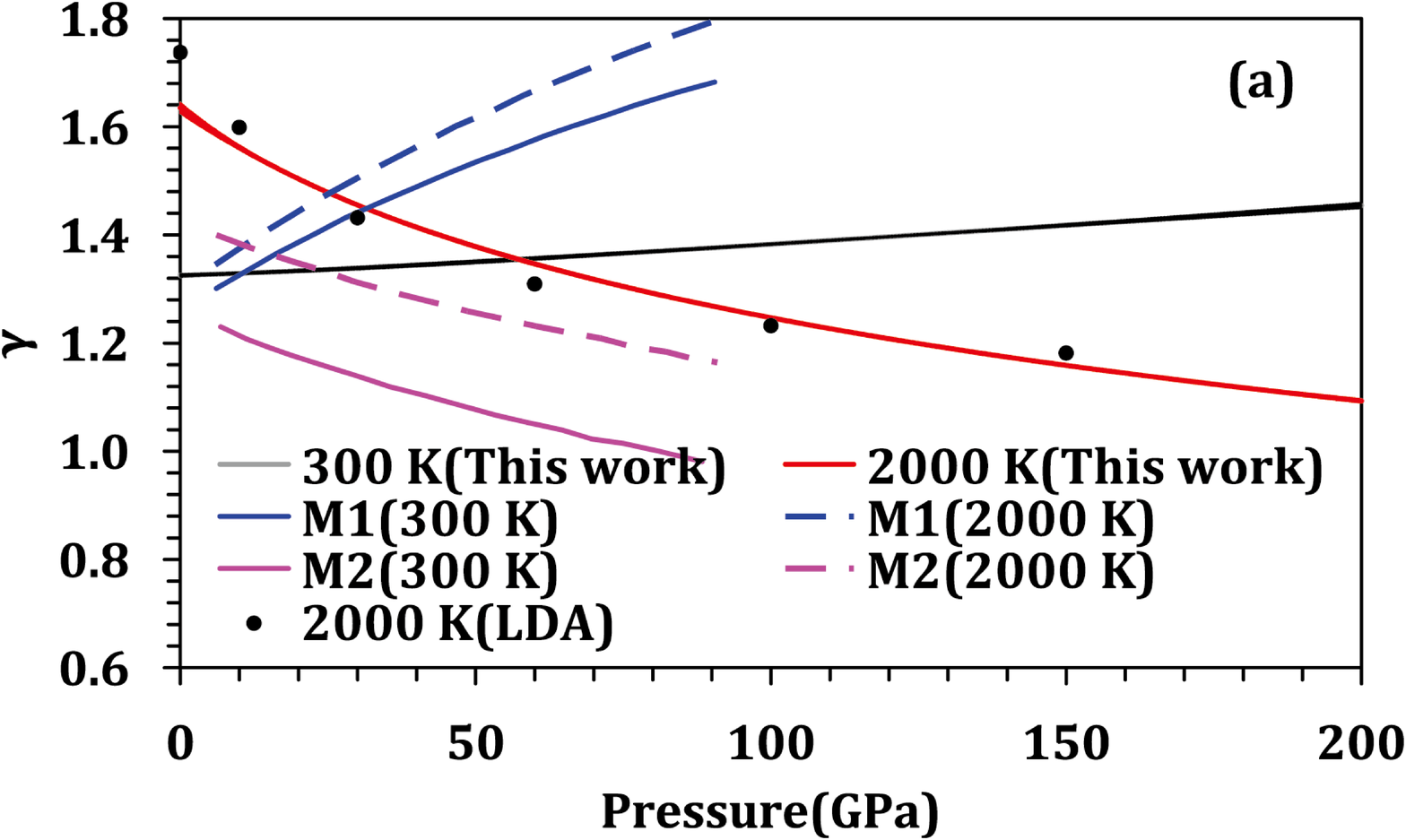}
\includegraphics[width=3.2in]{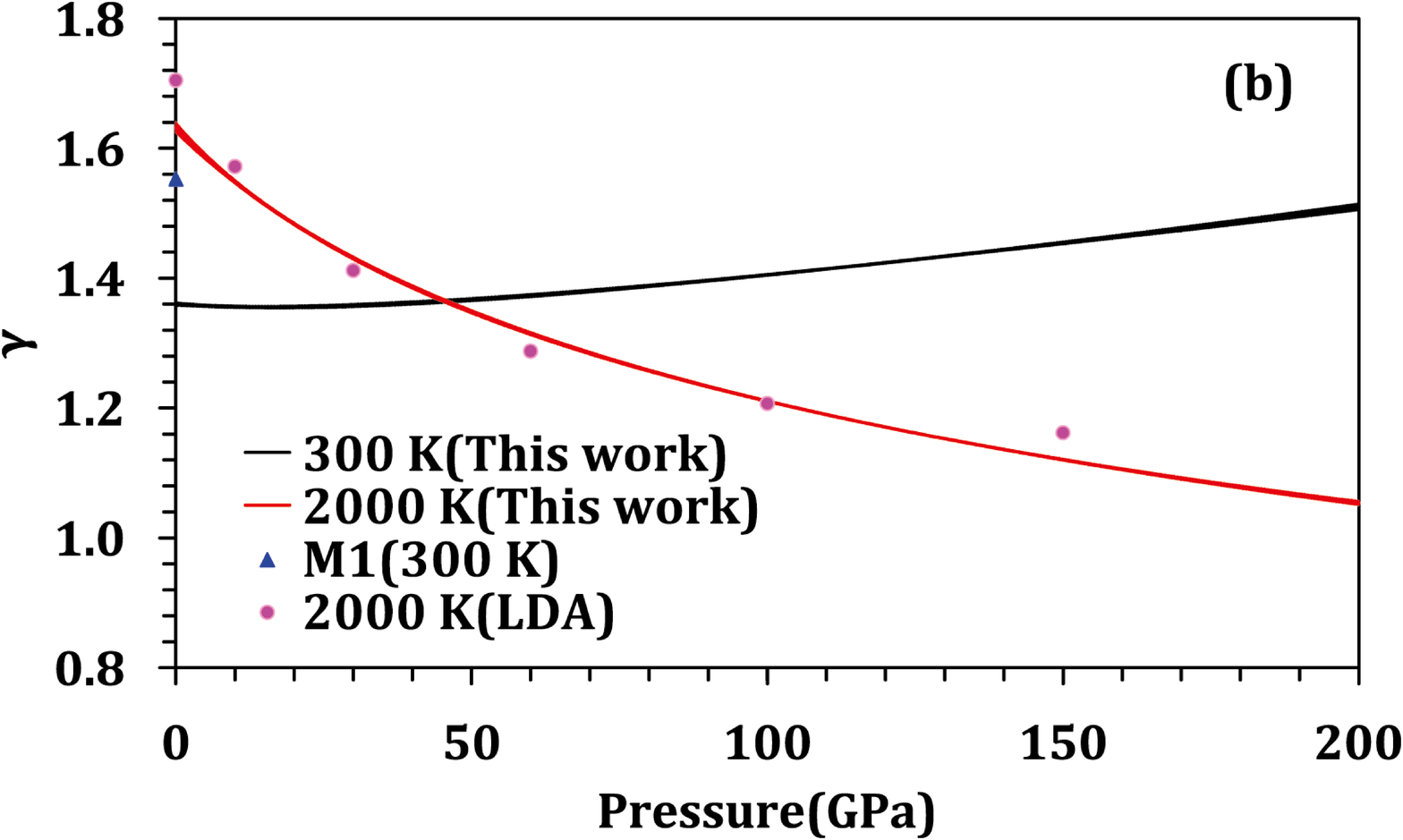}
\caption{Gr\"uneisen ratio as a function of pressure. (a) is for perovskite and (b) is for post-perovsksite. The shading of the curves indicate one-sigma width of the uncertainties. The references for all the models and DFT calculations are \cite{Gillet00}(a,M1 and M2), \cite{Tsuchiya05}{LDA in a and b}, and \cite{Guignot07}(b,M1).}
\label{fig:FiguresM4}
\end{figure}

The Gr\"uneisen ratio $\gamma$ is calculated by
\begin{equation}
\gamma=\frac{\alpha K_TV}{C_V}
\label{eq:Gamma}
\end{equation}
where $C_V$ is the constant volume heat capacity obtained from phonon calculations. The Gr\"uneisen ratios both for Pv and PPv fall in the region of previous models (Fig. \ref{fig:FiguresM4}).

\section{Conclusion}
We have presented QMC computations of $\text{MgSiO}_3$ equations of state and stability for both perovksite and post-perovskite. Our results showed that QMC not only gives good equations of state but also a reasonable Pv-PPv phase boundary for $\text{MgSiO}_3$ under lower mantle temperature conditions. For this iron free silicate, the predicted QMC Pv-PPv phase boundary may have a double-crossing of the geotherm, which would lead to a second Pv phase region just above the core mantle boundary. However, we could not conclude that this double-crossing will exist in the lower mantle, since the presence of Fe could change the Pv-PPv phase boundary dramatically \cite{Cohen05,Shieh06}.  The accuracy of QMC in this three component system has been demonstrated. It indicates that it is possible to further study the equations of state of the iron bearing silicate $\text{(Fe,Mg)SiO}_3$ using QMC simulations. 

\section{Acknowledgments}
\label{acknowledgments}
This work is supported by National Science Foundation grants DMS-1025370 and EAR-1214807. REC was supported by the Carnegie Institution and by the European Research Council advanced grant ToMCaT. This work used the Extreme Science and Engineering Discovery Environment (XSEDE) computers, which is supported by National Science Foundation grant number OCI1053575, and computers at the Oak Ridge Leadership Computing Facility at the Oak Ridge National Laboratory, which is supported by the Office of Science of the U. S. Department of Energy under Contract No. DE-AC05-00OR22725. An award of computer time was provided by the Innovative and Novel Computational Impact on Theory and Experiment (INCITE) program with Project CPH103geo. KD and BM acknowledge support under the U. S. Department of Energy under Contract No. DE-SC0010517. LS was Supported through Predictive Theory and Modeling for Materials and Chemical Science program by the Office of Basic Energy Science (BES), Department of Energy (DOE).  Sandia National Laboratories is a multiprogram laboratory managed and operated by Sandia Corporation, a wholly owned subsidiary of Lockheed Martin Corporation, for the U.S. Department of Energy's National Nuclear Security Administration under Contract No. DE-AC04-94AL85000. We thank Jane Robb for assistance with editing the manuscript.

\clearpage
\bibliographystyle{apsrev4-1}
\bibliography{MgSiO3-Pv-PPv}{}

\end{document}